\documentclass[twocolumn,superscriptaddress,nofootinbib,preprintnumbers,showpacs,tightenlines,notitlepage]{revtex4}
\usepackage[latin1]{inputenc}
\usepackage{graphicx}
\usepackage{amssymb}
\usepackage{color}
\usepackage{float}
\usepackage{amsmath}
\usepackage{amsfonts}
\usepackage{dcolumn}
\usepackage{hyperref}
\usepackage{amsthm}
\usepackage{color}
\usepackage{bm}

\def\3nab{\tilde{\nabla}}

\def\be {\begin{equation}}
\def\ee {\end{equation}}
\def\ba {\begin{eqnarray}}
\def\ea {\end{eqnarray}}

\newtheorem{lem}{Lemma}
\newcommand{\bra}[1]{\left(#1\right)}

\newcommand{\sfr}[2]{{\textstyle\frac{#1}{#2}}}

\newcommand{\E}{{\mathcal E}}

\newcommand{\barray}{\begin{array}}
\newcommand{\earray}{\end{array}}

\newcommand{\del}{\nabla}

\newcommand{\udot}{{\mathcal A}}

\begin{document}

\title{Gaussian curvature of spherical shells: A geometric measure of complexity}
\author{Sayuri Singh}
\email{sayurisingh22@gmail.com }
\affiliation{Astrophysics Research Centre, School of Mathematics, Statistics and Computer Science, University of KwaZulu-Natal, Private Bag X54001, Durban 4000, South Africa.}
\author{Dharmanand Baboolal}
 \email{Baboolald@ukzn.ac.za }
 \affiliation{Astrophysics Research Centre, School of Mathematics, Statistics and Computer Science, University of KwaZulu-Natal, Private Bag X54001, Durban 4000, South Africa.}
 \author{Rituparno Goswami}
\email{Goswami@ukzn.ac.za}
\affiliation{Astrophysics Research Centre, School of Mathematics, Statistics and Computer Science, University of KwaZulu-Natal, Private Bag X54001, Durban 4000, South Africa.}
\author{Sunil D. Maharaj}
\email{Maharaj@ukzn.ac.za}
\affiliation{Astrophysics Research Centre, School of Mathematics, Statistics and Computer Science, University of KwaZulu-Natal, Private Bag X54001, Durban 4000, South Africa.}

\begin{abstract}
 In this paper we consider a semitetrad covariant decomposition of spherically symmetric spacetimes and find a governing hyperbolic equation of the Gaussian curvature of two dimensional spherical shells, that emerges due to the decomposition. The restoration factor of this hyperbolic travelling wave equation allows us to construct a geometric measure of complexity. This measure depends critically on the Gaussian curvature, and we demonstrate this geometric connection to complexity for the first time. We illustrate the utility of this measure by classifying well known spherically symmetric metrics with different matter distributions. We also define an order structure on the set of all spherically symmetric spacetimes, according to their complexity and physical properties.
\end{abstract}
 
\pacs{04.20.-q, 04.40.Dg}
\maketitle
\section{Introduction}
The {\em complexity} of a physical system or a dynamical process is broadly defined in terms of the degree to which the components of the system engage in organised and structural interactions, via the underlying mathematical/logical  principles that govern the system or the process (see for example \cite{Com1,Com2,Com3,Com4,Herrera} and the references therein). It is widely believed that systems with very high complexity (under some suitable measure) may exhibit both randomness and regularity, and these may generate emergent phenomena. For example, a highly complex and sufficiently large quantum system gives rise to the classical properties (which can be broadly described as the emergence to the next level, as the mathematical principles governing the classical properties are quite different from those that govern the quantum properties). Although, the importance of the concept of complexity is widely accepted in scientific critiques of physical systems or processes, no general and accepted measures of complexity currently exists. This reflects our logical inability to build up an unified framework that cut across all the natural, social and artificial phenomena.\\

While this general measure has remained elusive, a large number of measures have been proposed for specific types of systems or processes, that produce the intuitively correct predictions at the extremities of the complexity spectrum. These new proposals are continuously being produced and tested across the interdisciplinary set of natural philosophers. One such class of systems are the self-gravitating systems that include the astrophysical and cosmological phenomena. It is widely accepted so far, that the underlying principles of these self-gravitating systems, obey the general theory of relativity, as developed by Einstein. As this theory is inherently geometrical in nature, study of the complexity of these systems will naturally generate some geometrical measure. 

\subsection{Previous works}

The importance of complexity in the context of general relativity was highlighted in the treatment of Herrera \cite{Herrera} when considering self-gravitating systems. In these systems, we can introduce a complexity factor that results from the orthogonal splitting of the Riemann curvature tensor. The complexity factor may be related to the inhomogeneity, the anisotropy and heat flux of dissipative systems. Therefore, the concept of complexity has been investigated in astrophysical bodies such as compact relativistic stars, neutron stars and radiating stars in general relativity. \\

Spherical symmetry, cylindrical symmetry and hyperbolic systems have been studied in various treatments \cite{Arias_Con_Fuen_Ram_2022, Cas_Con_Ov_Soto_Stuk_2019, Her_Di_Os_2018, Her_Di_Os_2019, Her_Di_Os_2020, Her_Di_Os_2021, Jas_Mau_Singh_Nag_2021, Mau_Gov_Kaur_Nag_2022, Mau_Nag_2022, Shar_B_2018, Shar_B2_2018, Shar_B_2019, Shar_Tar_2020}; this emphasizes the role of complexity in various geometries. It is interesting to note that issues related to complexity have been investigated in generalised gravity theories including Einstein-Gauss-Bonnet theory, the more general Lovelock theory, f(R) theory and other extended theories of gravity \cite{Abb_Naz_2018,Arias_Con_Fuen_Ram_2022,Shar_Maj_Nas_2019, You_2020, You_Bhatti_Has_2020, You_Bhatti_Nas_2020, Zub_Az_2020}. \\

Of particular interest are relativistic self-gravitating fluids and the influence of dissipative effects. There have been recent attempts to link the existence of first integrals of the motions in shear-free fluids, neutral and charged, to complexity \cite{Gum_Gov_Mah_2021,Gum_Gov_Mah_2022}. Also, the energy conditions for a composite matter distribution (combination of a viscous matter, null dust and null strings) exhibit complex behaviour \cite{Brass_Mah_Gos_2021}. Recent studies in complexity include gravitational collapse, embedding, gravitational coupling and static irrotational matter \cite{Bog_Gov_2022, Mau_Gov_Kaur_Nag_2022, Mau_Gov_Singh_Nag_2022, You_Bhatti_Khl_Asad_2022}, in radiating stars in general relativity.

\subsection{This paper}

The key aspect of most of these previous works, is that the background spacetime is specified, thereby restricting the definition of complexities within that given geometry. Hence, these definitions may not necessarily apply to all the elements of a given geometrical class (e.g. spherical symmetry). It is necessary, therefore, to study the problem under more general setting. For example, we must focus on the Locally Rotationally Symmetric (LRS-II) class of spacetimes if we wish to study spherically symmetric self-gravitating systems. This is exactly the approach in his paper. In order to reach a general definition for the geometrical measure of complexity, in this paper we start with the subclass of the gravitating systems that are spherically symmetric. \\

As described in the next two sections, this extra symmetry enables us to describe the system as a differential system of a set of covariantly defined geometrical and thermodynamical scalar variables, Note that any 4-dimensional spherically symmetric gravitating system can be viewed as the time evolution of 3-dimensional spaces which are foliations of 2-dimensional spherical shells. Hence the Gaussian curvature of these shells plays a crucial role in the overall dynamics of these systems. By explicitly constructing a second order hyperbolic wave equation, that obeys the causal structure of the spacetime and governs the evolution and propagation of the Gaussian curvature, we generate a measure of complexity. This measure not only satisfies the intuitive picture at the extremities, but enables us to impose an order relation on the set of spherically symmetric self-gravitating systems, and one can then generate a hierarchy of these systems in the ascending or descending order of their complexity, via this given measure.\\

The paper is organised as follows: In the next section we discuss the semitetrad decomposition of 4 dimensional spherically symmetric systems, using the timelike and preferred spacelike vectors. This decomposition naturally generates a set of geometric and thermodynamic scalars. Furthermore, the Ricci and Bianchi identities of the timelike and the preferred spacelike vectors give rise to the governing field equations in terms of these variables. In section III, we define the measure of complexity in terms of the hyperbolic wave equation governing the Gaussian curvature of the spherical shells. In section IV, we then use the definition to construct a classification of spacetimes according to their complexity. In section V, we define an order structure on the set of all spherically symmetric spacetimes, according to their complexity and physical properties.

\section{Spherically symmetric spacetimes in 1+1+2 Covariant formalism}

We know that the timelike unit vector ${u^{a}}$ ${\left(u^{a}\, u_{a} = -1\right)}$ can be used to split the spacetime locally in the form ${\mathcal{R} \otimes \mathcal{V}}$, where ${\mathcal{R}}$ is the timeline along ${u^{a}}$ and ${\mathcal{V}}$ is the 3-space perpendicular to ${u^{a}}$ \cite{Covariant}. Thus the metric becomes
\begin{equation}
 g_{ab} = - u_{a}\,u_{b}+ h_{ab},
\end{equation}
where ${h_{ab}}$ is the metric on 3-space perpendicular to $u^a$. The covariant time derivative along the observers' worldlines, denoted by `${^{\cdot}}$', is defined using the vector ${u^{a}}$, as
\begin{equation}\label{dot}
\dot{Z}^{a ... b}{}_{c ... d} = u^{e}\,\nabla_{e}\, Z^{a ... b}{}_{c ... d},
\end{equation} 
for any tensor ${Z^{a...b}{}_{c...d}}$. The fully orthogonally projected covariant spatial derivative, denoted by `${D}$' , is defined using the spatial projection tensor ${h_{ab}}$, as
\begin{equation}\label{D}
D_{e}\,Z^{a...b}{}_{c...d} = h^r{}_{e}\,h^p{}_{c}\,...\, h^q{}_{d}\,h^a{}_{f}\,...\, h^b{}_{g}\,\nabla_{r}\,Z^{f...g}{}_{p...q},
\end{equation}
with total projection on all the free indices. The covariant derivative of the 4-velocity vector ${u^{a}}$ is decomposed irreducibly as follows
\begin{equation}
\nabla_{a}\, u_{b} =  -u_{a}\,A_{b}+D_a\,u_b= -u_{a}\,A_{b} + \frac{1}{3}h_{ab}\,\Theta + \sigma_{ab}\,,
\end{equation}
where ${A_{b}}$ is the acceleration, ${\Theta}$ is the expansion of ${u_{a}}$, ${\sigma_{ab}}$ is the shear tensor. Note that spherical symmetry dictates that the tensor `$D_a\,u_b$' is symmetric and hence the rotation term becomes identically zero. Furthermore the energy momentum tensor of matter, decomposed relative to ${u^{a}}$, is given by
\begin{eqnarray} \label{3.Tab}
T_{ab} &=& \mu\, u_{a}\,u_{b} + p\, h_{ab} + q_{a}\,u_{b} + u_{a}\, q_{b} + \pi_{ab},
\end{eqnarray}
where ${\mu}$ is the effective energy density, ${p}$ is the isotropic pressure, ${q_{a}}$ is the 3-vector defining the heat flux and ${\pi_{ab}}$ is the anisotropic stress. Further to this, the Weyl tensor is only described by it's traceless symmetric electric part (the magnetic part is identically zero in the absence of rotation)
\begin{eqnarray}
E_{ab} &=& C_{abcd}\,u^{c}\, u^{d}.
\end{eqnarray}
The spacetimes that are spherically symmetric have an important property: they exhibit
locally (at each point) a unique preferred spatial direction, covariantly defined, which creates a local axis of symmetry. We denote the unit vector along this preferred spatial direction as $e^a\, (e^ae_a=1\,,\,e^au_a=0)$. Thus the 3-space ${\mathcal{V}}$ is now further split by this vector $e^a$, and the 1+1+2 covariantly decomposed spacetime is given by \cite{chris1}
\begin{equation}\label{2.Nab}
g_{ab} = -u_{a}\,u_{b} + e_{a}\, e_{b} + N_{ab},
\end{equation}
where ${N_{ab}}$ ${\left(e^{a}\, N_{ab} = 0 = u^{a}\, N_{ab}, N^{a}{}_{a} = 2\right)}$ projects vectors onto 2-spaces called `2-sheets' , orthogonal to ${u^{a}}$ and ${e^{a}}$, which in this case are the surfaces of spherical shells. We introduce a new derivatives for any tensor ${\Phi_{a...b}{}^{c...d}}$:
\begin{eqnarray}
\label{hatderiv}
\hat{\Phi}_{a...b}{}^{c...d} &\equiv& e^{f}\,D_{f}\, \Phi_{a...b}{}^{c...d}\,.
\end{eqnarray}
The  1+3 kinematical and Weyl quantities and anisotropic fluid variables are split irreducibly as
\begin{eqnarray}
A^{a} &=& \mathcal{A}\,e^{a},\\
\sigma_{ab} &=& \Sigma\left(e_{a}\, e_{b} - \frac{1}{2}\, N_{ab}\right) , \\
\label{2.qa} q_{a} &=& Q\,e_{a}, \\
\label{2.pia} \pi_{ab} &=& \Pi\left(e_{a}\, e_{b} - \frac{1}{2}\,N_{ab}\right),\\
E_{ab} &=& \mathcal{E} \left(e_{a}\, e_{b} - \frac{1}{2}\, N_{ab}\right).
\end{eqnarray}
The fully projected 3-derivative of ${e^{a}}$ is given by
\begin{eqnarray}
D_{a}\,e_{b} &=&\frac{1}{2}\,\phi\,N_{ab}\,,
\end{eqnarray}
when traveling along ${e^{a}}$, ${\phi}$ is the sheet expansion. 

\subsection{The field equations}

Thus the set of quantities that fully describe the spherically symmetric class of spacetimes are $\{\udot, \Theta, \phi, \Sigma, \E, \mu, p, \Pi, Q\}$. 
The propagation and evolution equations for the LRS-II variables can be obtained using the Ricci identities of the vectors $u^a$ and $e^a$, along with the doubly contracted Bianchi identities (see \cite{Clarkson_2007} for more details). \\

\textit{Propagation}:
\ba
\hat\phi  &=&-\sfr12\phi^2+\bra{\sfr13\Theta+\Sigma}\bra{\sfr23\Theta-\Sigma}
    \nonumber\\&&-\sfr23\bra{\mu+\Lambda}
    -\E -\sfr12\Pi,\,\label{phihat}
\\
\hat\Sigma-\sfr23\hat\Theta&=&-\sfr32\phi\Sigma-Q, \label{Sigthetahat}
 \\
 \hat\E-\sfr13\hat\mu+\sfr12\hat\Pi&=&
    -\sfr32\phi\bra{\E+\sfr12\Pi}+\bra{\sfr12\Sigma-\sfr13\Theta}Q . \label{Ehatmupi}
\ea
\smallskip

\textit{Evolution}:
\ba
   \dot\phi &=& -\bra{\Sigma-\sfr23\Theta}\bra{\udot-\sfr12\phi}+Q, \label{phidot}
\\
\dot\Sigma-\sfr23\dot\Theta&=&-\udot\phi +2(\sfr13\Theta-\sfr12\Sigma)^2 
\nonumber \\ && +\sfr13(\mu+3p-2\Lambda)-\E+\sfr12\Pi, \label{Sigthetadott}
 \\
 \dot\E-\sfr13\dot\mu+\sfr12\dot\Pi&=& (\sfr32\Sigma-\Theta)\E+\sfr14(\Sigma-\sfr23\Theta)\Pi
   \nonumber \\ &&+\sfr12\phi Q -\sfr12(\mu+p)(\Sigma-\sfr23\Theta). \label{Edotmupi}
\ea
\smallskip

\textit{Propagation/evolution}:
\ba
   \hat\udot-\dot\Theta&=&-\bra{\udot+\phi}\udot+\sfr13\Theta^2
    +\sfr32\Sigma^2 \nonumber\\
    &&+\sfr12\bra{\mu+3p-2\Lambda}\ ,\label{Raychaudhuri}
\\
    \dot\mu+\hat Q&=&-\Theta\bra{\mu+p}-\bra{\phi+2\udot}Q \nonumber \\
&&- \sfr32\Sigma\Pi,\,
\\    \label{Qhat}
\dot Q+\hat
p+\hat\Pi&=&-\bra{\sfr32\phi+\udot}\Pi-\bra{\sfr43\Theta+\Sigma} Q\nonumber\\
    &&-\bra{\mu+p}\udot\, .
\ea 
\\
The 3-Ricci scalar of the spacelike 3-space orthogonal to $u^a$ can be expressed as
\be ^3 R = -2[\hat\phi+\sfr34\phi ^2-K],  \ee
\\
where $K$ is the Gaussian curvature of the two dimensional shells, defined by $^2R_{ab} = KN_{ab}$. We can write the Gaussian curvature $K$ in terms of the covariant scalars as
\be K=\sfr13\mu-\E-\sfr12\Pi+\sfr14\phi ^2 -(\sfr13\Theta-\sfr12\Sigma)^2. \label{k} \ee
\\
The evolution and propagation equations for the Gaussian curvature $K$ are given by
\ba 
\dot K &=& (\Sigma-\sfr23\Theta-)K, \label{kdot} \\ 
\hat K &=& -\phi K. \label{khat}
\ea
\subsection {Misner-Sharp mass for spherically symmetric spacetimes}
In terms of the 1+1+2 scalars defined above, we can derive the Misner-Sharp \cite{Misner_Sharp_1964} mass equation for spherically symmetric spacetimes, that describes the mass enclosed within a spherical shell at a given instant of time. This is given by \cite{Stephani}
\be \label{mass}
\mathcal M (r,t) = \frac{C}{2}(1-\del_aC\del^aC), 
\ee
where $C$ is the physical radius (the area radius) of the spherical shell under consideration. We can then write $C = \frac{1}{\sqrt{K}}$ , where $K$ is the Gaussian curvature of the spherical two dimensional shells. Hence, the mass can be expressed as
\be \label{mass1}
\mathcal M = \frac{1}{2\sqrt{K}}\left(1-\frac{1}{4K^3}\del_aK\del^aK\right). 
\ee
Using the equations (\ref{k}), (\ref{kdot}) and (\ref{khat}), the Misner-Sharp mass now takes the form
\be 
\mathcal M = \frac{1}{2K^{3/2}}\left[\sfr13\mu-\E-\sfr12\Pi\right] \,.
\label{Misner} 
\ee 
Using the field equations and equation (\ref{Misner}), the variations of the Misner-Sharp mass along $u^a$ and along the preferred spatial direction $e^a$ are given as 
\ba 
\mathcal {\hat M} &=& \frac{1}{4K^{3/2}}\left[ \phi\mu-(\Sigma-\sfr23\Theta)Q \right] \,, \\
\mathcal {\dot M} &=& \frac{1}{4K^{3/2}}\left[(\Sigma-\sfr23\Theta)(p+\Pi)-\phi Q \right] \,.
\ea
It is interesting to note here that both the matter terms (the energy density and anisotropic pressure) as well the source free gravity term (the electric part of the Weyl scalar) contribute to the Misner-Sharp mass in the spherically symmetric system. This was the reason for this term to have primary importance in describing the complexity of gravitating system by Herrera \cite{Herrera}. However, in this paper, we explore a step further to provide a geometrical order relation for the complexity as will be evident in the next few sections. 

\section {Gaussian curvature of 2-shells and complexity}

From the structure of the evolution and propagation equations (\ref{kdot}) and (\ref{khat}) of the Gaussian curvature of two dimensional spherical shells, it is very obvious that one can combine them together to write a closed form wave equation. We argue, that the structure of this wave equation is intimately related to the underlying complexity of the self-gravitating system. We first clarify the following points:
\begin{enumerate}
\item {\bf Why Gaussian curvature?} From the covariantly decomposed geometry of the spherically symmetric spacetimes, it is clear that every three dimensional constant time slice can be represented as the foliations of these spherical two dimensional shells. Furthermore, the time evolution of these foliations would then determine the complete four dimensional geometry. Hence the manner in which these foliations evolve depending on other geometrical and thermodynamic variables, will definitely determine how complex the system would be. 
\item {\bf Why wave equation?} From the structure of the evolution and propagation equations of the Gaussian curvature, it is clear that they can be combined into any second order equation. However, the information about the change in the Gaussian curvature of a given 2-shell, must travel causally in the spacetime. Hence, the underlying second order equation, that determines this causal behaviour, must be hyperbolic in nature. This naturally rules out the elliptic equations. Also what makes this more interesting, is that combining these two first order equations we obtain the wave equation in a source-free and closed form, which is much simpler to analyse. 
\end{enumerate}

To derive the second order wave equation that determines how information about any change in the Gaussian curvature of a given 2-shell propagates in the spacetime, let us define the following variables
 \ba
\Psi &\equiv& \Sigma - \sfr23\Theta, \label{Psi} \\
X &\equiv& \E-\sfr13\mu+\sfr12\Pi, 
\ea
In terms of these new variables, equations (\ref{phihat}), (\ref{Sigthetadott}), (\ref{k}) and (\ref{Misner}) become
\ba
\hat\phi &=& -\sfr12\phi^2-\Psi^2-\Psi\Theta-X-\mu, \label{phihat2}\\
\dot\Psi &=& \sfr12\Psi^2-\udot\phi-X+p+\Pi , \label{Psidot}\\
K &=& -X+\sfr14(\phi^2-\Psi^2), \label{k2}\\
\mathcal M &=& \frac{-X}{2K^{3/2}} \label{Misner1}.
\ea
Now taking the dot derivative and hat derivative of (\ref{kdot}) and (\ref{khat}) respectively, we get
\ba 
\ddot K &=& \dot\Psi K + \Psi^2 K, \label{kddot} \\ 
\hat {\hat K} &=& -\hat\phi K +\phi^2 K. \label{khhat}
\ea
Subtracting the second equation from the first, we get the required closed form wave equation
\be 
\ddot K - \hat{\hat K} +\mathcal{F} K =0\,,
\ee   
where we have defined 
\be
\mathcal{F} =-[\dot\Psi+\Psi^2+\hat\phi-\phi^2]\,.
\label{F1}
\ee
This term `$\mathcal F$' mimics the {\em restoration factor} of a travelling disturbance in an elastic medium. Now using (\ref{Psi}), (\ref{phihat2}) and (\ref{Psidot}), we have 
\ba 
\dot\Psi+\Psi^2+\hat\phi-\phi^2&=&-\sfr32(\phi^2-\Psi^2)-\Psi^2-\Psi\Theta \\ \nonumber
&&-\udot\phi-2X-\mu+p+\Pi. 
\ea
Since from (\ref{k2}) we have $\phi^2-\Psi^2=4(K+X)$, we can substitute in the above to get the general form of the restoration factor as
\be
\mathcal{F}=\left[6K-16\mathcal M K^{3/2}+\Psi^2 +\Psi\Theta+\udot\phi+\mu-p-\Pi\right].
\label{F2}
\ee
 It is interesting to see that in the most general scenario, the restoration factor depends (either explicitly, or via the equation of state) on all the geometrical and thermodynamical variables, which form the set
 \be
 \mathcal{D}\equiv\{\udot, \Theta, \phi, \Sigma, \E, \mu, p, \Pi, Q\}\,.
 \ee
 Now we are in a position to define the complexity of a given spherically symmetric spacetime by the following lemmas:
 \begin{lem}
The complexity of a spherically symmetric self-gravitating system, is completely determined by the independent 1+1+2 geometrical and thermodynamical variables.
\end{lem}
\begin{lem}
The complexity of a spherically self-gravitating system necessarily depends on the Gaussian curvature of 2-dimensional spherical shells, that emerges due to the 1+1+2 decomposition. 
\end{lem}
\begin{lem}
The number of independent  1+1+2 geometrical and thermodynamical variables, that the restoration factor arising from the travelling wave equation of Gaussian curvature of 2-spherical shells depends on, 
gives a measure of the complexity of the spherical self-gravitating system.
 \end{lem}
 
 \section{A classification of spacetimes according of their complexity}
 
From the definition of complexity of spherically symmetric self-gravitating systems, as stated in the previous section, we can now construct a hierarchy of spacetimes according to their complexity as follows:
\begin{enumerate}
\item \textbf{Minkowski spacetime:} This is undoubtedly the case of least complexity. We have $\mathcal M=\Psi=\udot=\mu=p=\Pi=0$. Therefore, the restoration factor is given by 
\be
\mathcal{F}=6K. 
\ee
From the equation (\ref{k}), it is evident that the restoration factor only depends on the single variable $\phi\in\mathcal{D}$.

\item \textbf{Rindler spacetime}: Spherically symmetric Rindler spacetime is the Minkowski spacetime as conceived by an accerelated observer in the preferred spatial direction. As the accelration scalar $\udot$ is non-zero in this case, we have 
\be
\mathcal{F}=6K+ \udot\phi,
\ee
 and the restoration factor depends on the subset $\{\phi, \udot\}\subset\mathcal{D}$.
 
\item \textbf{Schwarzschild spacetime, exterior to the horizon:} As we know, this is a vacuum spacetime in presence of a central point mass. Exterior to the horizon, the spacetime is static and we have $\mathcal M= \mathcal M_0,$ which is a spacetime constant. Furthermore we have $\Psi=\mu=p=\Pi=0$ and $-\udot\phi=\E=-2\mathcal M_0 K^{3/2}.\,$ Hence the restoration factor is given as
\be 
\mathcal{F}=[6K-14\mathcal M_0 K^{3/2}]. 
\ee
Using the field equations it is clear that the restoration factors depends on the subset $\{\phi, \E\}\subset\mathcal{D}$.

\item \textbf{Schwarzschild spacetime, interior to the horizon:} This spacetime is non-static but spatially homogeneous. Again we have $\mathcal M= \mathcal M_0,$ and $\phi=\mu=p=\Pi=0$ and $\E=-\Psi(\Psi+\Theta)\,.$ Hence the restoration factor is given as
\be 
\mathcal{F}=[6K-14\mathcal M_0 K^{3/2}-\E]. 
\ee
Using the field equations we see that restoration factor depends on the subset $\{\Theta, \E\}\subset\mathcal{D}$.

\item \textbf{Reissner-Nordstrom spacetime:} This is spherically symmetric electro-vac spacetime in the presence of a central charge. In this case the Misner Sharp mass $\mathcal M$ is a function of the curve parameter of the integral curves of the preferred spacelike vector $e^a$. Further to this, we have $\mu=3p=\sfr32\Pi$ and $\Psi=0$. Although the form of the restoring factor is similar to the Schwarzschild case, we have an extra dependence of $\mu$ via the Gaussian curvature $K$ in $\mathcal F$. Hence the restoration factor depends on the subset $\{\phi, \E, \mu\}\subset\mathcal{D}$.

\item \textbf {Spherical static stars:} Starting from the Schwarzschild interior solution, numerous classes of static spherically symmetric stellar models have been obtained, as the solution to the field equations. In general, as these models are static, we have $\Theta=\Sigma=Q=0$ and the mass function depends on the curve parameter of the integral curves of $e^a$ only. In these cases the restoration factor is generally given as
\be
\mathcal{F}=\left[6K-16\mathcal M K^{3/2}+\udot\phi+\mu-p-\Pi\right],
 \ee
 and it depends on the subset $\{\phi, \udot, \E, \mu, p, \Pi\}\subset\mathcal{D}$.

\item \textbf{Vaidya spacetime:} This is the spacetime consisting of unpolarised radiation outside a radiating star. Here the mass function depends on the curve parameters of null vector $u^a+e^a$. For this spacetime, we have  $\mu=Q=3p=\sfr32\Pi, \; \Sigma=\sfr23\Theta$ and  $\udot\phi=\mu-\E=\mu+2\mathcal MK^{3/2}.$ Therefore, the restoration factor is given by
\be 
\mathcal F=[6K-14\mathcal M K^{3/2}+Q]\,.
\ee
Hence the restoration factor depends on the subset $\{\phi, \E, Q\}\subset\mathcal{D}$.

\item \textbf{Friedmann Lemaitre Robertson Walker spacetime:} This is the spacetime containing homogeneous and isotropic dust. In this case we have $\udot=p=\Pi=\E=0$ and $\Psi^2+\Psi\Theta=-\sfr29\Theta^2$ (since $\Psi=-\sfr23\Theta$). We also have $\mu=6\mathcal M K^{3/2}$. Also  the Friedmann constraint (from the field equations) relates $\mu$ and $\Theta$. Therefore, the restoration factor is given by
\be 
\mathcal F=[6K-10\mathcal M K^{3/2}-\sfr29\Theta^2], 
\ee
which depends of the subset $\{\phi, \Theta\}\subset\mathcal{D}$.

\item \textbf{Kantowski Sachs dust universe:} This spacetime describes spatially homogeneous dust, in presence of non-trivial shear and Weyl scalars. In this case we have $\phi=\udot=p=\Pi=\E=0$ and $\E+\sfr23 \mu= -\Psi(\Psi+\Theta)$. Hence the restoration factor is given by
\be 
\mathcal{F}=[6K-14\mathcal M K^{3/2}-\E-\sfr43 \mu]. 
\ee
 Using the field equations we see that restoration factors depends on the subset $\{\Theta, \E, \mu\}\subset\mathcal{D}$.
 
\item \textbf{Lemaitre Tolman Bondi dust spacetime:} This is the spacetime describing spherically symmetric inhomogeneous dust. In this case we have $\udot=p=\Pi=0$ and $\Psi^2+\Psi\Theta=(\Sigma-\sfr23\Theta)(\Sigma+\sfr13\Theta)$. Therefore the restoring factor is given by
\ba
\mathcal F&=& [6K-16\mathcal M K^{3/2}\nonumber\\
&&+(\Sigma-\sfr23\Theta)(\Sigma+\sfr13\Theta)+\mu] ,
\ea
and the factor depends on the subset $\{\phi, \E, \Theta, \Sigma, \mu\}\subset\mathcal{D}$.

\item \textbf{Generalised Vaidya spacetime:} This is a spacetime that contains a specific mixture of Type I and Type II matter fields. Here the mass function $\mathcal M$ is a function of both the integral curves of the null vector $(u^a+e^a)$ and $e^a$. In this case we also have $\Sigma=\sfr23\Theta$. Therefore the restoration factor is given by
\be
\mathcal F= [6K-16\mathcal M K^{3/2}+\udot\phi+\mu-p-\Pi],
\ee
and depends on the subset $\{\phi, \E, \Theta, \mu, Q, p, \Pi\}\subset\mathcal{D}$,  We note the dependence on $Q$ is implicit here via the equation of state of the Type I matter field. 

\item \textbf{General spherical radiating stars:} These spacetimes describe general spherically symmetric combination of Type I and Type II matter fields, This is the other extremity, where $\mathcal F$ depends on all the variables in $\mathcal D$ and this is the most complex spherically symmetric self-gravitating system.
\end{enumerate}

The spacetimes listed above are a sample of metrics which arise in practice in applications in cosmology and astrophysics. Clearly lemmas 1-3 may be considered for other spacetime geometries that are spherically symmetric. We point out that composite distributions with Type I and Type II fluids have been shown to be relevant for general spherical radiating stars \cite{BB1,BB2,BB3}. These hold in four and higher dimensions and the matter distribution satisfy generalised energy conditions as demonstrated by \cite{Brass_Mah_Gos_2021,BB4}.

\section{A physical order relation on spherically symmetric spacetimes}

From the Definition 1, and the examples given in previous section, it is clear that the set of all spherically symmetric spacetimes can be ordered according to their complexity. This can be done in a number of ways. The most straightforward approach would be to count the number of independent geometrical and thermodynamical variables in the restoration factor for the spacetimes and order them accordingly. This makes the POSET (partially ordered set) of the spherically symmetric spactimes a {\em chain} where every elements are comparable to each other.\\

This approach, however doesn't consider the physical properties of the spacetimes. For example, in this approach the complexity of the Schwarzschild spacetime is exactly the same for regions outside and inside the horizon. However, physically, the region outside the horizon is quite different from the region inside the horizon. The former is static while the later is non-static and spatially homogeneous. Similarly, in this picture Vaidya spacetimes and Kantowski Sachs spacetime have the same order of complexity, while physically these two spacetimes have very different properties. \\

To incorporate the physical properties of the spacetimes, along with their respective complexities,  we define a new order relation on the set of all spherically symmetric self gravitating system $\mathcal S$. From the discussion in the previous section, it is clear that one can uniquely define a mapping 
\be
\Phi: \mathcal S\rightarrow \mathcal{P}(\mathcal D)\setminus \{\}\,,
\ee
(where $\mathcal{P}(\mathcal D)$ is the power set of $\mathcal D$), by considering the subset of $\mathcal D$ on which the restoration factor of a given spacetime depends on. Now we know that the set 
$\mathcal{P}(\mathcal D)\setminus \{\}$ carries a natural set inclusion order `$\le$', defined as
follows
\be
A, B \in \mathcal{P}(\mathcal D)\setminus \{\}\;\;,\;\; A\le B \;\;iff \;\;A \subseteq B\;.
\ee
Hence, by virtue of the mapping $\Phi$, one can make the set $\mathcal S$ carry the same set inclusion order relation. The key features of this order relation are as follows:
\begin{enumerate}
\item From the definition of the given order relation, it is obvious that not all spactimes are comparable to each other. 
\item Also within the subset of comparable spacetimes, a transparent picture of how a given spacetime can be physically generalised to obtain a spacetime with higher complexity, emerges. 
\end{enumerate}
 Fig 1., describes a subset of  $\mathcal S$, under this set inclusion order.
\begin{figure}[h!!!!!] 
   \centering
   \includegraphics[width=7cm]{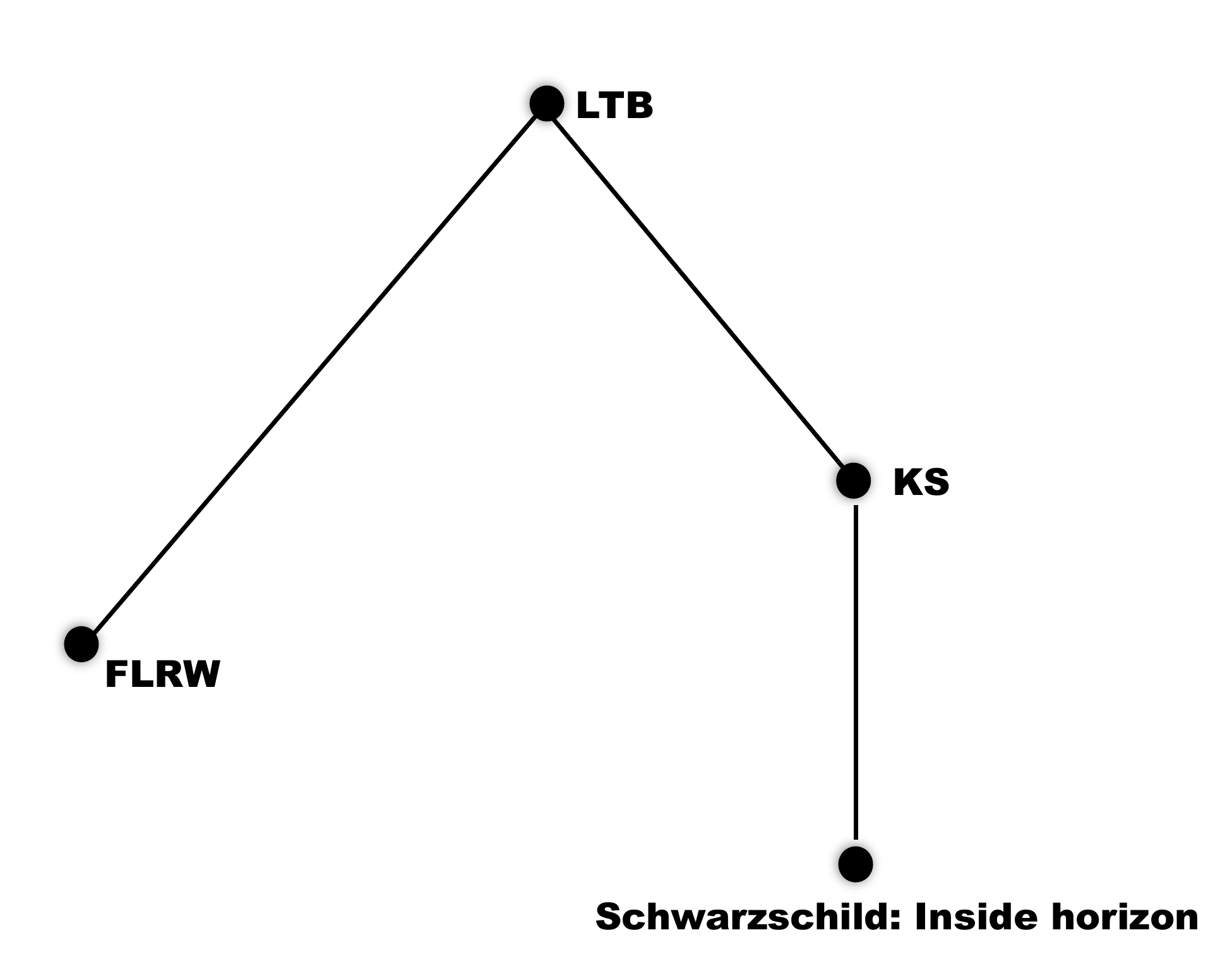} 
\caption{Hasse diagram according to the order relation, relating some well known dust/vacuum spacetimes.}
\end{figure}
From the definition of the set inclusion order, it is clear that the set $\mathcal S$ has a maximum element, which describes the class of most general spherically symmetric spacetimes with a general combination of Type I and Type II matter fields. However, $\mathcal S$ contains no minimum element (rather a set of {\it minimal elements}, depicting spherically symmetric spacetimes with less complexities). 
Also since $\mathcal{P}(\mathcal D)\setminus \{\}$, is a  {\it join semilattice} under the set inclusion order (every pair of elements have a least upper bound in the set), we can also claim that $\mathcal S$ is also a join semilattice, where the join or supremum of any two spacetimes exists (which describes the spacetime with the least upper bound of complexity for these given spacetimes).

\section{Discussion}

In this paper, we performed a semitetrad covariant 1+1+2 decomposition of spherically symmetric spacetimes, where the 2-spaces are spherical shells. We used the governing second order wave equation for the Gaussian curvature of these spherical shells to construct a transparent measure of the complexity of these spacetimes. Our results are summarised in lemmas 1-3 connecting geometrical and thermodynamical variables, Gaussian curvature and complexity. We clearly demonstrated by taking examples of several well known classes of spherically symmetric spacetimes, how this well defined measure can be implemented. Further to this, taking into account this new measure as well the inherent physical and geometrical properties of different spacetimes, we constructed a well-defined order relation on the set of all spherically symmetric spacetimes to show that this set becomes a join semilattice under this order. \\

This is indeed a novel approach, erstwhile unexplored, and the natural way forward would be to include more general self-gravitating system within the sphere of consideration. For example the general Locally Rotationally symmetric self-gravitating systems and beyond, where the semitetrad decomposition would yield geometrical quantities that can be linked with the measures for complexities in those spacetimes. We keep this for future work.

\begin{acknowledgments}
SS. RG, DB and SDM are supported by National Research Foundation (NRF), South Africa. 
\end{acknowledgments}

\end{document}